%% file: main.tex
\documentclass[submission,copyright,creativecommons]{eptcs}
\providecommand{\event}{MARS 2017} 
\usepackage{breakurl}             
\usepackage{underscore}           

\usepackage{graphicx}
\usepackage{fancyvrb}
\usepackage{latexsym}
\usepackage{amssymb}
\usepackage{float}
\usepackage{multicol}
\floatstyle{ruled}
\newfloat{program}{thp}{lop}
\floatname{program}{Table}
\title{Modelling, Verification, and Comparative Performance Analysis of the B.A.T.M.A.N. Protocol}

\author{Kaylash Chaudhary  \institute{\parbox[c][5em][c]{0.275\textwidth}{\centering \em School of Computing, Information, and Mathematical Sciences\\  University of the South Pacific}}
\and
Ansgar Fehnker  \institute{\parbox[c][5em][c]{0.275\textwidth}{\centering \em Formal Methods and Tools\\
University Twente}}\and Vinay Mehta \institute{\parbox[c][5em][c]{0.275\textwidth}{\centering \em School of Computing, Information, and Mathematical Sciences\\  University of the South Pacific}}
}

\def\titlerunning{Modelling and Verification of the B.A.T.M.A.N. Protocol}
\def\authorrunning{K.~Chaudhary, A.~Fehnker and V.~Mehta}

\newcommand{\batman}{\textsc{B.A.T.M.A.N.}}
\begin{document}

\maketitle
\begin{abstract}
This paper considers on a network routing protocol known as Better Approach to Mobile Ad hoc Networks (\batman{}). The protocol serves two aims: first, to discover all bidirectional links, and second, to identify the best-next-hop for every other node in the network. A key element is that each node will flood the network at regular intervals with so-called originator messages.

This paper describes in detail a formalisation of the \batman{} protocol. This exercise revealed several ambiguities and inconsistencies in the RFC. We developed two models. The first implements, if possible, a literal reading of the RFC, while the second model tries to be closer to the underlying concepts. The alternative model is in some places less restrictive, and rebroadcasts more often when it helps route discovery, and will on the other hand drop more messages that might interfere with the process.

We verify for a basic untimed model that both interpretations ensure loop-freedom, bidirectional link discovery, and route-discovery. We use simulation of a timed model to compare the performance and found that both models are comparable when it comes to the time and number of messages needed for discovering routes. However, the alternative model identifies a significantly lower number of suboptimal routes, and thus improves on the literal interpretation of the RFC.\end{abstract}

\input{intro.tex}

\input{batman.tex}

\input{datastructures.tex}
\input{handling.tex}
\input{processing.tex}

\input{uppaal.tex}

\bibliographystyle{eptcs}
\bibliography{reference}

\end{document}

%% file: intro.tex
\section{Introduction}\label{sec:intro}
Since the introduction of wireless mobile
ad hoc networks, many protocols have been used for making
communication efficient and by finding the best possible routes
from source to destination. The German ``Freifunk'' community developed the network routing protocol known as Better Approach to Mobile Adhoc Network \batman{}  as an alternative to OLSR. \batman{} is a proactive protocol, for detecting all bidirectional links and identifying the best-next-hop for all other nodes. The protocol is table driven and maintains route information and updates by keeping information on received originator messages in a so-called \emph{sliding window}. A node records how many messages from another node with a sequence number in a given range of recent sequence numbers were received via which neighbor.

This paper describes a model based on the RFC \cite{BatmanRFC}. During the implementation of the protocol as an Uppaal model it turned out that a number of conditions were not unambiguously defined in the context of looping sequence number and a limited local view of a node. The formalisation will define those terms. The RFC also contained a few conditions that were inconsistent.  To address this we developed two models: The literal interpretation of the RFC implements the protocol with a reading that is as close as possible to the text of the RFC. The alternative interpretation resolves inconsistencies in light of the underlying concepts of \batman{}. For example, a literal reading of the description will exclude certain messages from  being recorded in the sliding window, while the general description of the sliding window suggests that those messages should be recorded.

For both interpretations we verify a few basic properties for an untimed model and a small four node topology. This is mostly meant to help with debugging the models, and ensure e.g. that the bidirectional link check works as expected. This analysis show that both, the literal and alternative model, meet this minimum standard. We then use a timed model in a 17 node network to analyse the performance, in particular the quality of the chosen ``best-next-hops''. These are sometimes suboptimal and actually not the best next hop. This analysis shows that the alternative model reduces the occurrences of suboptimal ``best-next-hops''.

The performance of \batman{} in comparison to OLSR has been analysed in a real environment in \cite{stairs,hallway}. A simulation of a real environment has been used to analyse \batman{} in \cite{paderborn,batman4omnet}. Furlan simulated the performance of \batman{} for a select type of network topologies \cite{furlan2011analysis}. These studies consider a particular implementation with specific hardware in a specific environment and do not analyse the routing algorithm in isolation. A formal analysis of \batman{} was conducted by Cigno and Furlan in \cite{cignoimproving}. This analysis discovered that routing loops are possible, and proposed improvements that ensure loop freedom. This work was complemented by simulation studies and real measurements. While this work does not use model checking per se, it presents a formal model to study routing loops. In comparison, this paper presents a formal Uppaal model, and uses verification and simulation to study the quality of route discovery. Uppaal has been used perviously to study other protocols, such as LUNAR, OLSR, AODV, DYMO, and LMAC \cite{bulychev2012uppaal,fehnker2013topology,fehnker2012automated,Kamali2015,wibling2004automatized}

This paper is organised as follows. The next section describes the \batman{} routing protocol and in particular the sliding window. Section \ref{sec:sliding} introduces the common data structures, Section \ref{sec:handling} and \ref{sec:processing} the handling and processing of OGMs. These section give rise to the two alterative models. Section \ref{sec:untimed} and \ref{sec:timed} describe the templates and the verification and simulation results, respectively.

%% file: batman.tex
\section{The \batman{} protocol}\label{sec:batman}
The major objective of a network routing protocol is to discover routes available in a network.  An active user group known as the German ``Freifunk'' community has been involved in the development of a network routing protocol known as Better Approach to Mobile Adhoc Network \cite{BatmanRFC}. The RFC abbreviates the protocol as \batman{}, including the dots. The protocol is meant as an alternative to OLSR (Optimized Link State Routing Protocol), a proactive wireless routing protocol widely used in mobile ad hoc networks \cite{OLSR}. Instead of having each node compute routing tables that capture the complete network topology as OLSR does, \batman{} nodes maintain only information on the neighboring nodes. This information is used to identify the best-next-hop \cite{Huhtonen04comparingaodv}. It is then a desired property of \batman{} that the best-next-hop will indeed realize the best route.

To maintain a fresh list of available neighbours, and to deal with changing topologies, nodes will send \emph{originator messages} (OGMs) at regular intervals. OGMs are forwarded under certain circumstances to neighbouring nodes, first to correctly identify all available bidirectional links. Once these have been established a node selects a neighbour as best-next-hop to a destination, based on the number of OGMs from that destination that have passed through that neighbor within a given time frame, called the \emph{sliding window}. The protocol drives on OGMs flooding the network and is phrased in terms of the origin of the OGMs. Once routes have established along bidirectional links, these originators will be the destination for packets to be send.

The phases of finding bidirectional links and finding the best-next-hop are however not clearly divided. They may overlap for different nodes, and nodes will iterate through these phases continually, as the sliding window moves, and bidirectional links expire.

\subsection*{Terminology}
The following terminology is used by the \batman{} routing protocol\cite{BatmanRFC}.

\begin{itemize}
\item \textbf{OGM}. \emph{Originator messages}. A message send periodically by a node to announce the existence of a node. Each OGM contains the following information:
    \begin{itemize}
    \item The \emph{OID} (Originator ID). Once an OGM is created, the OID will remain unchanged, even when it is rebroadcasted.
    \item The \emph{SID} (Sender ID). A node that rebroadcasts an OGM will set the SID to its own ID.
    \item A \emph{sequence number}. According to the RFC it has a range $0$ to $2^{16}-1$. The sequence number will be incremented if a node generates a new OGM. Once it exceeds the upper bound, it will start at $0$ again.
    \item The TTL (Time-to-live).  The TTL decremented each time an OGM is rebroadcasted by another node.  Once it reaches the value 1, the OGM will be dropped. The TTL is also a measure for the distance travelled.
    \item A \emph{uni-directional link flag}. Used to indicate that an OGM was received from a link that is presumed to be uni-directional.
    \item A \emph{direct-link-flag}. Used when a node rebroadcasts an OGM  where the sender was also the originator.
    \end{itemize}
\item \textbf{Routing Table} Each node maintains a routing table with the following information:
    \begin{itemize}
    \item \textbf{Bidirectional Sequence Number} The sequence number of the last self-originating OGM that was rebroadcasted by a neighbor.
    \item \textbf{Sliding Window}
    The \emph{sliding window} specifies for each originator a range of recent sequence numbers. OGMs with sequence number in this range will be recorded. A neighbour which rebroadcasted the most OGMs for an originator within the sliding window is considered the \emph{best-next-hop} for that originator.
    \item
    For each originator the \textbf{last TTL} and the \textbf{last sequence number}.
    \end{itemize}
\end{itemize}

\subsection*{Example}
\begin{figure}[t!]
\includegraphics[width=\linewidth]{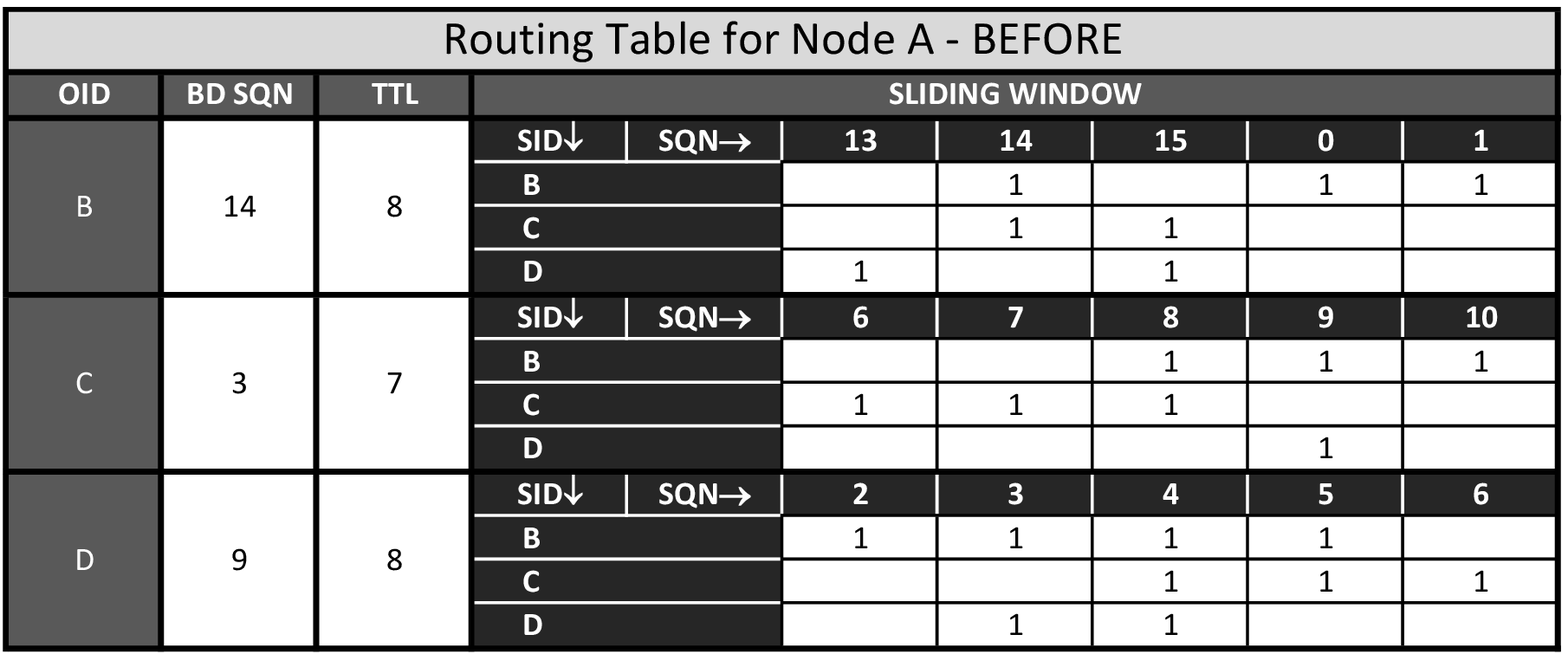}
\includegraphics[width=\linewidth]{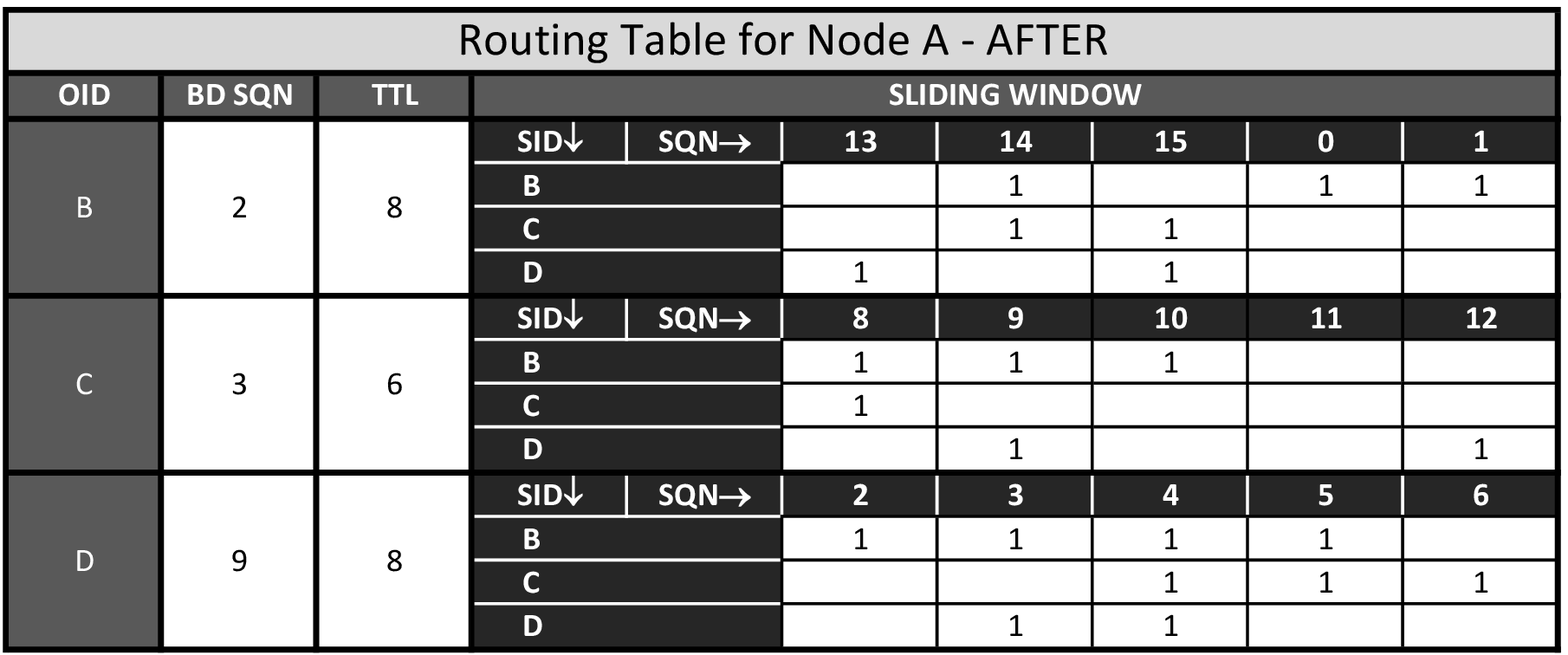}\vspace{-2ex}
\caption{Routing table for A before and after processing two OGMs from originators B and C.}\label{fig:routing}
\end{figure}

Figure \ref{fig:routing} gives a example routing table for node A in for a small network with four nodes. The bidirectional sequence number for originator B is in this example is 14, and the last TTL is 8. The sliding window for originator B encompasses sequence numbers 13 to 1. Within the window, A received an OGM from originator B, three times from Node B, twice from Node C and twice from Node D. This means that Node B is currently the best-next-hop for Node B.

Similarly we have for originator C -- according to this example -- that both Node B and Node C have broadcasted three times an OGM from C within the window, and Node D only once. This means that both B and C are considered to be the best-next-hop for originator C.

Suppose that Node A will receives an OGM rebroadcasted by Node B, with OID A,  sequence number 2, and the direct-link-flag set to true. In this case the OGM is an echo of an OGM node A sent itself. From this follows that there exists a bidirectional link. The bidirectional link sequence number (BD SQN) for OID B will be changed to 2.

Suppose further that A receives an OGM, rebroadcasted by Node D, with OID C, and sequence number 12. The range of the sliding window for C was sequence number 6 to 10. After reception, the range of the sliding window will be 8 to 12. Data of sequence numbers 6 and 7 will be dropped, and D will be recorded as having sent an OGM with sequence number 12. After this update node B will have rebroadcasted 3 out the last 5 OGMs, Node C one, and Node D two. Node B will be the presumed best-next-hop for originator C. See Figure \ref{fig:routing} for the new routing table after these updates.

%% file: datastructures.tex
\section{Data Structures and Constants}\label{sec:sliding}
This section describes the common data structures and data structures used by of the different Uppaal model. The subsequent section will describe the processes and conditions for handling OGMs.

Constant are the number of nodes \texttt{N}, the maximum  sequence number \texttt{MAX\_SQN}, the initial TTL \texttt{TTL\_MAX}, and the size of the sliding window \texttt{WINDOW\_SIZE}. We considered models with up to 25 nodes, a maximum sequence number of 15, an initial TTL of up to 10, and a window size between 2 and 8. The maximum sequence number of $15$ is a lot smaller than the recommended value of $2^{16}-1 (=65535)$. The reason is that we want to analyse the behaviour of the sliding window, including the behaviour when sequence numbers loop.

Datatypes have been defined for sequence numbers, node IDs, TTLs, and indices into arrays of the sliding window. Node IDs take values from $0$ to $N-1$, and we reserve the node ID $-1$ for an invalid ID; it is used as padding where we had to use fixed sized arrays to implement variable sized buffers.

The datatype \texttt{OGM} in Table \ref{tab:datatypes} is used for the OGMs. Nodes communicate by writing an OGM to a shared global variable \texttt{ogmglobal}, and synchronizing on a channel from array \texttt{broadcast chan SendOGM[N]}. Each node maintains a buffer of received, but not yet processed OGMs, \texttt{ogmlocal}, which is implemented as a fixed array of type \texttt{OGM}. The next OGM to be processed is \texttt{ogmlocal[0]}. The size of this buffer is chosen such to avoid buffer overflow -- we use values up to 64. A node increments counter \texttt{buffererror} if a buffer overflow occurs.

Each node maintains an array \texttt{table} of size \texttt{N} of type \texttt{OriginatorInfo}. Each entry contains the bidirectional sequence number \texttt{biDirectionalSqn}, i.e. the last sequence number received from the originator \texttt{sqn}, the time-to-live \texttt{ttl}, and the sliding window \texttt{window}. The latter is an array with an boolean array of size \texttt{WINDOW\_SIZE} for each node in the network. The last element in this array, refers to sequence number \texttt{sqn}, the second last element to \texttt{sqn-1}, etc.  The model includes a number of functions to update the window for a given \texttt{oid} consistently, e.g. to move the entries, when the sequence increases. Of course this takes into account that the sequence number may loop.

\begin{program}[t]
\begin{multicols}{2}
\begin{verbatim}
typedef struct {
   anyIdT oid;
   anyIdT sid;
   ttlT ttl;
   sqnT sqn;
   bool isDirect;
   bool isUnidirectional;
} OGM;

typedef struct {   	
   bool entries[WINDOW_SIZE];
} SlidingWindow;

typedef struct {
   sqnT biDirectionalSqn;
   sqnT sqn;
   ttlT ttl;
   SlidingWindow window[N];
} OriginatorInfo;
\end{verbatim}

\end{multicols}\vspace*{-2ex} \caption{Datatype definitions for OGMs and routing table entries. }\label{tab:datatypes}
\end{program} 

%% file: handling.tex
\section{Handling of Originator Messages}\label{sec:handling}
This section gives a detailed account of how the model relates to the \batman{} RFC. A key to model the \batman{} protocol is to faithfully represent Section 5.2 of the RFC. It describes how to classify different types of OGMs. This paper uses the numbering scheme from the RFC to refer to these rules.\\[0.5ex]

\noindent\textbf{{Rule 5.2.1}} This rule states to drop OGMs sent with a different version of the protocol. It is not included in the model, since we assume that all nodes run the same version.

\noindent\textbf{{Rule 5.2.2}} This rule states to drop OGMs a node receives from its own transmitter. Since the automaton modelling a node cannot synchronize with itself, this scenario is excluded by the semantics of synchronisation in Uppaal. Note, this rule has nothing to do with dropping messages that are a rebroadcast of a message that has been sent earlier by this node. That is dealt with elsewhere.

\noindent\textbf{{Rule 5.2.3}} This rule deals with nodes that have multiple transmission interfaces. It is not included, since the model only considers nodes with a single interface.

\noindent\textbf{{Rule 5.2.4}} This rule deals with the bidirectional-link check. It is modeled by boolean function \texttt{bool meetsRule5\_2\_4(OGM ogm)}. It returns true if \texttt{ogm.oid} is equals to the nodes own ID. In this case the bidirectional link information may have to be updated, as described in the next section.

\noindent\textbf{{Rule 5.2.5}} This rule deals also with the bidirectional-link check. The OGM is dropped if the uni-directional-link flag \texttt{ogm.isUnidirectional} is set. This rule is modelled by boolean function  \texttt{bool meetsRule5\_2\_5(OGM ogm)}.

\noindent\textbf{{Rule 5.2.6}} This rule determines which OGMs should be used to update the sliding window. This update is referred to as \emph{Process 5.4}, and will be explained later. \emph{Rule 5.2.6} states to invoke \emph{Process 5.4} if the OGM is received via a bidirectional link and contains, as the RFC state a ``New Sequence Number (is NOT a duplicate)''.

    This apparently simple statement requires some interpretation, since the RFC does not define how to determine whether a sequence number is new. The model has to assume a local view on the network, and cannot refer to a global notion of ``duplicate'' or ``new''.  We also have to take into account that sequence numbers may loop, and that OGMs may not arrive in the order they have been created.

    Given an OGM from originator \texttt{ogm.oid} and sequence number \texttt{ogm.sqn}, and the most recently recorded sequence number \texttt{table[oid].sqn}. We modeled ``new sequence number'' to mean that the value of \texttt{ogm.sqn} is between the \texttt{table[ogm.oid].sqn} and \texttt{table[ogm.oid].sqn} plus half of the range of the sequence numbers, modulo the range.

    The comment ``NOT a duplicate'' in brackets, might seem like a clarification, but complicates things further. Being not a duplicate is not necessarily the same as having a new sequence number. Within the context of the sliding window, a message with a sequence number in the range of the sliding window is not a duplicate, if for the same \texttt{ogm.oid} no OGM from the same sender \texttt{ogm.sid} with same sequence number \texttt{ogm.sqn} was received. These OGMs are not duplicates, but also not newer.

    Recall that this rule is the precondition for \emph{Process 5.4}. This process however clearly states that it should only use OGMs with a newer sequence number. It does not mention duplicates.

    We decided to create two different models that explore the different alternatives. The first model applies a \emph{literal interpretation}, i.e. \emph{Rule 5.2.6} only applies if an OGM has a newer sequence number.  This model implements the condition that was explicitly stated in the description of \emph{Process 5.4}.

    This interpretation, however seems to undermine the purpose of the sliding window, namely to collect information on OGMs with sequence numbers in the range of the sliding window. The \emph{alternative interpretation} of \emph{Rules 5.2.6} is to consider OEMs with a newer sequence number, or OGMs that are in range but not duplicates.

    This rule is modelled by boolean function  \texttt{bool meetsRule5\_2\_5(OGM ogm)}.

\noindent\textbf{{Rule 5.2.7}} This rules deals with which OGMs will be rebroadcasted. Before rebroadcast it has to execute \emph{Process 5.5} which will be described later. The first set of OGMs that have to be rebroadcasted, are those that have been received from a single hop neighbor, i.e. for which \texttt{ogm.oid} equals \texttt{ogm.sid}. This includes, but not exclusively, the OGMs that take part in the bidirectional link check.

In addition OGMs that satisfy the following will be rebroadcasted according to the RFC:
\begin{quote} \em The OGM was received via a Bidirectional link AND via the Best
          Link AND is either not a duplicate or has the same TTL as the
          last packet which was not a duplicate (last TTL).
\end{quote}
The ``bidirectional link'' is the link between the receiving node and the sender of the OGM, \texttt{ogm.sid}. The specific condition for whether a node considers a link to be bidirectional will be discussed later.

To be the ``Best Link'' means that the sender \texttt{ogm.sid}  is a best-next-hop for originator \texttt{ogm.oid}. This decision is based on the sliding window. We will discuss later how to determine whether a node is a best-next-hop.

To determine whether an OGM is a duplicate we use the same interpretation as for \emph{Rule 5.2.6}. The sequence number is newer, or in range, and no OGM with the same \texttt{ogm.sid}, \texttt{ogm.oid}, and \texttt{ogm.sqn} has been received previously.

The last TTL is  the value of \texttt{table[oid].ttl}. The short summary here does not quite align with the description later in the RFC that describes \emph{Process 5.5}. We will discuss this in the next section. Also, while it is not explicitly mentioned in the RFC, the condition on the TTL can only be reasonably applied if the sequence number of OGM is in range. OGMs that are neither newer, nor in range should be dropped regardless of the TTL.

Note furthermore that this only tests for equality of the TTL of the OGM, with the last TTL. It disregards OGMs with an improved TTL, i.e. OGMs that arrived via a shorter route. This seems to contradict the purpose of the sliding window, namely to collect information to determine the best hops, and thus the best route. In contrast, the condition that the OGM should not be a duplicate does not consider the TTL at all. It means that OGMs will be rebroadcasted that are not a duplicates but have lower TTLs, i.e. OGMs that arrived via longer route.


Also here we decided to consider the alternatives. Both interpretations require that the link to the sender is bidirectional and received via a best-next-hop. The literal interpretation will require in addition that the sequence number is newer, or that the OGM is in range but not a duplicate, or in range and that its TTL equals the last TTL.

The alternative interpretation requires that OGM is newer, or that the OGM is in range \textbf{and} not a duplicate \textbf{and} also has TTL that is equal or better than the last TTL.

Finally, recall that \emph{Rule 5.2.7} is a precondition for \emph{Process 5.5}. The description of this process requires explicitly  that the TTL is at least 2, and to drop it otherwise. We include this in \emph{Rule 5.2.7}.

This complete rule is modelled by boolean function  \texttt{bool meetsRule5\_2\_7(OGM ogm)}.

It should be noted that these rules, in particular \emph{Rule 5.2.6} and \emph{Rule 5.2.7} are not mutually exclusive. Either of those rules may apply to an OGM or both. The RFC does not specify explicitly what to if the latter is the case. To ensure correct computation of the best-next-hop it is necessary that OGMs that satisfy both rules, are used to update the sliding window, and are rebroadcasted.

It should also be noted that these rules are not complete; there can be OGMs to which none of these rules apply. The RFC does not mention explicitly what to do; it appears that these OGMs have to be silently dropped.

%% file: processing.tex
\section{Processing of Originator Messages}\label{sec:processing}
The RFC describes three procedures to process OGMs, one for the bidirectional-link-check, one to update the neighbor ranking, and the last one to update an OGM before rebroadcast.

\noindent\textbf{{Process 5.3: Bidirectional Link Check}} This process checks if an OGM is a recent self-originating message, i.e. if \texttt{ogm.sid} is it own ID, the \texttt{ogm.isDirect} flag  set, and the sequence number \texttt{ogm.sqn} equal to its current number \texttt{sqnNode}. If so it stores \texttt{ogm.sqn} in  \texttt{table[oid].biDirectionalSqn}.

    The description of \emph{Process 5.3} also defines when a link is considered bidirectional: if the difference between the current sequence number \texttt{sqnNode} and the stored \texttt{table[oid].biDirectionalSqn} is less than some threshold \texttt{BI\_LINK\_TIMEOUT}.  Unfortunately, the RFC does not specify values for \texttt{BI\_LINK\_TIMEOUT}; for the model we use values between 2 and 10 (half the size of the sliding window, and up to twice its size).


\noindent\textbf{{Process 5.4: Neighbor Ranking}}
  This process updates the sliding window. The update adds a new  entry for sequence number \texttt{ogm.sqn} and originator \texttt{ogm.oid}. If the sequence number \texttt{ogm.sqn} is newer than the last recorded sequence number \texttt{table[oid].sqn}, the sliding window will be shifted before that update. This means that the oldest entries will be deleted and new entries up to sequence number \texttt{ogm.sqn} added.

  Note, that this process also defines a condition for updating the sliding window, even though \emph{Rule 5.2.6} serves the same purpose. However, these rules are not the same. Here, the precondition is that an OGM has have a newer sequence number. As mentioned before, the literal model uses this condition as \emph{Rule 5.2.6}. The \emph{alternative model} relies on the alternative interpretation of \emph{Rule 5.2.6}.

  Note furthermore, that the RFC mentions that the \emph{packet count} must be updated. However, a \emph{packet count} has not been introduced nor used elsewhere in the RFC. We assume that it is used as synonym for the neighbor ranking based on the sliding window.

  Finally, the RFC states that each node has to nominate one node from the set of nodes with the most OGMs in range to be the unique best-next-hop. This is how we implemented it for the literal model. In the alternative we leave this choice open; if two or more nodes happen to have the same ranking, they will all be considered a next-best-hop.


  \noindent\textbf{{Process 5.5: Rebroadcasting}}
  An OGM has to be rebroadcasted when it satisfies \emph{Rule 5.2.7}. Before rebroadcast the following fields of the OGM will be changed:
  \begin{itemize}\item The TTL \texttt{ogm.ttl} has to be decremented by one.
  \item If the sender is equal to the originator, i.e. if \texttt{ogm.sid} equals \texttt{ogm.oid}, then the \texttt{isDirect} flag will be set to true. This allows the originator - a direct neighbor - to distinguish between OGMs that were received directly from that neighbor, or via a detour.  Note, that the RFC also mentions as requirement that the OGM will sent the OGM back to the sender/originator. Since our model uses a only broadcast, this is satisfied by default.
  \item The \texttt{isUniDirectional} flag has to be set to true, if the  link to the sender \texttt{ogm.sid} does not satisfy the bidirectional-link requirement.
  \end{itemize}

Note furthermore, that the RFC mentions that OGMs with a TTL of 1 (or 0 after decrementing) have to be dropped. In the model we incorporated this in Rule 5.2.7, which effectively drops the message.


%% file: uppaal.tex
\section{Untimed Model}\label{sec:untimed}

The untimed model serves to assure that the bidirectional link check works, and that routes are discovered, and whether routes are loop-free for a static topology. It uses the same template for each node \texttt{id}.

The template has only two control locations. The first, is a committed initial location,  the second, labelled \texttt{Processing}, models normal processing of OGMs. The edge from the initial location to location \texttt{processing} initialises the sliding windows and other local variables.

From location \texttt{Processing} there are 8 self loops. They  model sending and receiving of OGMs, and the processing of the OGMs:
\begin{itemize}
\item An edge that synchronizes on channel \texttt{SendOGM[oid]!}. This transaction creates an OGM, and copies it to the shared global variable \texttt{ogmGlobal}.
\item An edge that synchronizes on channel \texttt{SendOGM[sid]?}, where \texttt{sid} is a valid sender ID. The guard \texttt{topology[sid]} ensures that sender and receiver are connected.  This transaction calls a function \texttt{receiveOGM()}, which copies the OGM from global variable \texttt{ogmGlobal}, and appends it to the local buffer \texttt{ogmLocal} unless it satisfies \emph{Rule 5.2.5}.
    \item Five edges that model the processing of OGMs in the buffer. One for when \emph{Rule 5.2.4} applies, one for when \emph{Rule 5.2.6} applies but not \emph{Rule 5.2.7}, one for when  \emph{Rule 5.2.7} applies but not \emph{Rule 5.2.6}, one for when both apply, and one for when the buffer is not empty, and neither rule applies. If Rule 5.2.7 applies, the edge will synchronize on channel \texttt{SendOGM[oid]!}, i.e. it forwards an OGM. All others are labelled with broadcast channel \texttt{tau[id]}. These edges model a local update of the routing information. The label allows use to give those transition a higher priority -- to reduce the number of states -- while it will not affect the routes that will be discovered.
\end{itemize}

The model contains a few auxiliary functions to add and remove OGMs to the buffer, to  manage the sliding window, and to process the OGMs. It also includes a number of function that are used to specify properties, e.g. \texttt{bool hasLoop()} which is used to check for loop freedom, or function \texttt{int countBidirectionalMisses()} which counts how many bidirectional links still need to be discovered.

\begin{figure}[t!]\begin{centering}{\includegraphics[width=\linewidth]{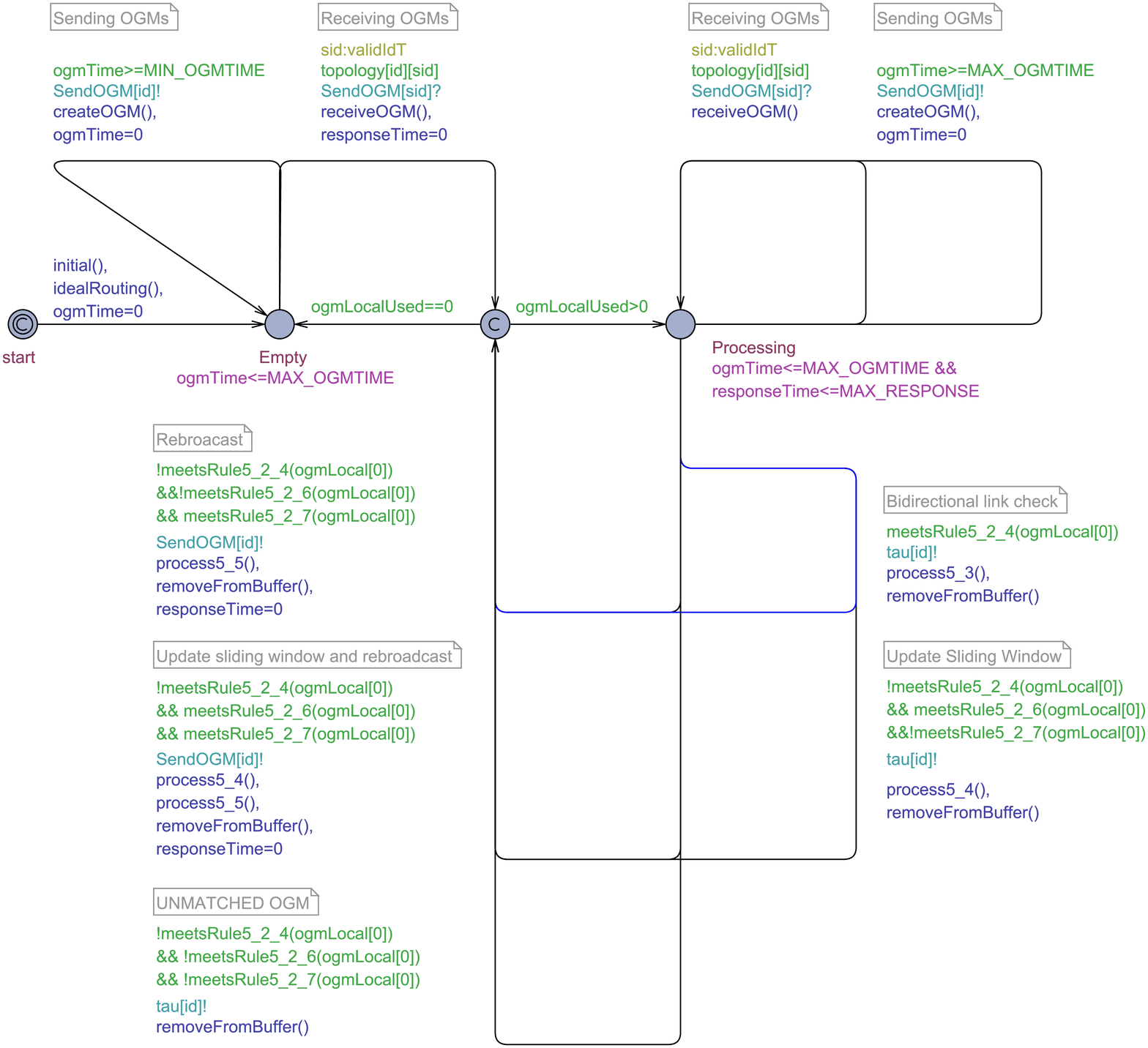}}\caption{Timed model for simulation}\label{fig:model}\end{centering}
\end{figure}

\paragraph{Verification Results}
For the verification we limit ourself to a model with four nodes in a ring. Each nodes sends one OGM, and node 0 a second one. For this we check the following:
\begin{itemize}
\item \texttt{A[] !hasloop()}. Whether the routes have loops.
\item \texttt{A[] (forall(id:validIdT) Node(id).OGMremain==0 \&\& sumBuffer() ==0 )\\\hspace*{2em} imply countBidirectionalMisses() == 0} \\This checks that once all OGMs have been sent and processed each node other than 0 have discovered a route all bidirectional links have been discovered (no misses).
\item \texttt{A[] ((forall(id:validIdT) Node(id).OGMremain==0\&\& sumBuffer() ==0 ) \\\hspace*{2em} imply  (forall(id:validIdT) (id==0 || Node(id).table[0].bestNext>-1))}\\ This checks that once all OGMs have been sent and processed, each node other than 0 have discovered a route to node 0.
\end{itemize}
The model satisfies all of these. It should be noted that we show absence of loops for a small static network with only 4 nodes. In general, sequence number based protocols have been shown to not to be loop free \cite{van2013sequence}, and  \cite{cignoimproving} presents a specific scenario for \batman{} with a changing topology.

The literal and alternative models behave are except for the implementation of the functions implementing \emph{Rule 5.2.4} to \emph{Rule 5.2.7} and  \emph{Processes 5.3} to \emph{Process 5.5}, as discussed previously. There is only one notable difference in the data structure for the routing table. The literal model also includes a field \texttt{table[oid].nextBest}, since the literal interpretation requires a unique best-next hop for each originator.

Verification was performed on an Intel i5-5200 CPU 2.2Ghz processor with 8 GB RAM running Uppaal 4.1.19. Verification of these properties took 19 seconds for either model. Verifying these same properties for a 5 node network, with one extra central node, runs out of memory after about 600 seconds.

\section{Timed Models}\label{sec:timed}

The timed model uses simulation to analyse and compare the performance of the two versions of the protocol; in particular how quickly the bi-directional link check occurs, how quickly routes are discovered, and the quality of the routes.

The timed models uses the same data-structures and functions as the untimed model, but they attach time bounds on when events can occur. The model uses clock \texttt{ogmTime} to ensure that any node sends an OGM once between \texttt{MIN\_OGMTIME} and \texttt{MAX\_OGMTIME}. The transition that model creation and sending of new OGMs includes  guard \texttt{ogmTime>=MIN\_OGMTIME}, and all control location invariant \texttt{ogmTime<=MAX\_OGMTIME}.

The model uses clock \texttt{responseTime} to ensure that a node rebroadcasts within \texttt{MAX\_RESPONSE} time units, while there is an OGM in the buffer. The model includes two control locations, \texttt{Empty} for when the buffer is empty, \texttt{Processing} for when the node processes OGMs. In the latter location there is the additional invariant \texttt{responseTime <=MAX\_RESPONSE}.

All other transitions model internal updates and are labelled with urgent broadcast channel \texttt{tau[id]!}. This means these transition are taken as soon as they are enabled. See Figure \ref{fig:model} for a depiction of node template.

The model also includes a number of auxiliary functions to support analysis of the performance by simulation, such as function \texttt{countRouteMismatch()} which counts how many ``best-next-hops'' are actually  suboptimal.

\begin{figure}[t!]\includegraphics[width=\linewidth]{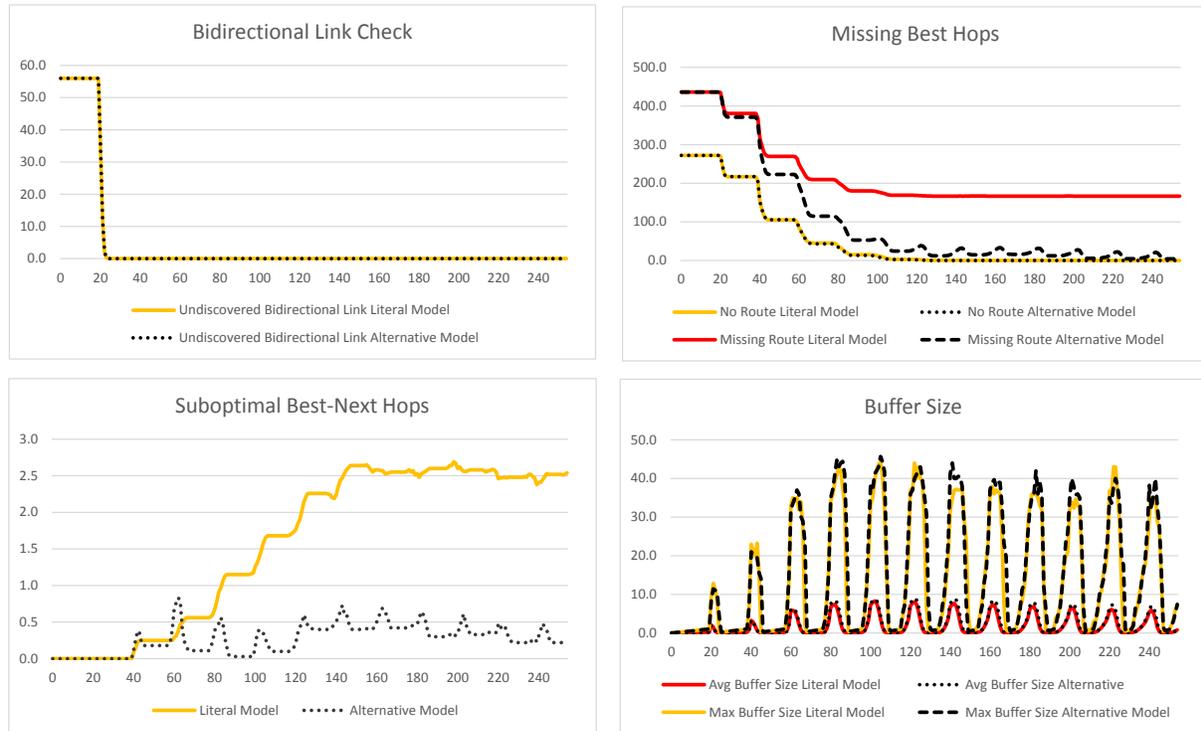}
\caption{Simulation results. The results are averages over 100 runs, except for the maximum buffer size, which is the maximum over 100 runs.}\label{fig:simulationresults}
\end{figure}
\paragraph{Simulation Results. }

For the simulation we consider a 4 by 4 grid with 16 nodes, plus one additional node in the center. The one additional node is connected to the four central nodes of the grid. The purpose of the central node is to introduce irregularity to the grid. The model furthermore assumes that processing of OGMs takes at most 1 time unit. New OGMs are created once between 19 and 20 time units.  The buffer for each node has maximum size 64. For these setting we ran a simulation of 100 runs up to time $255$, which means it consider 12 rounds of OGMs being sent and processed. The results are depicted in Figure \ref{fig:simulationresults}. Simulation of one run takes about 320 seconds for the literal model, and 330 second for the alternative model. The presented results are for 100 simulations each.

The number of undetected bidirectional links drops to zero after the first round of OGMs have been sent. There is no significant difference between the literal and alternative model, as the changes do not affect the bidirectional link check.

The results for the number of routes for which no ``Best-next-hop'' declines within 6 rounds to zero. This corresponds to the 6 hops of longest route in a 4 by 4 grid; from the one corner to the diagonally opposite corner. There is no significant difference between the two model in this respect.

There is a noticeable with respect to the number of potential best-next-hops. The literal model selects a designated best-next-hop. This has consequences further on, since a node only forwards OGMs that were received via a best-next-hop. The literal model will rebroadcast only OGM received via the designated best-next-hop. The alternative will rebroadcast OGMs received by any best-next-hop.

This then has further consequences when we consider the number of route errors. These are nodes that have been identified as best-next-hop, while they are not. In the literal model the number of routing errors at time 250, when the 12th round of OGMs has been processed, is 2.5.  In the alternative model this value is only 0.22.  It appears $85\%$ of all simulations of the literal model have at least one routing error at time 250, while this is true for only $17\%$ of all runs in the alternative model. Since the alternative message rebroadcasts more OGMs from more best-next-hops, the chances that a suboptimal route wins the neighbor ranking is greatly diminished.

It should also be seen in light that a suboptimal ``best-next-hop'' in the literal model is the designated ``best-next-hop''. A node has no alternatives. The alternative model identifies all potential best-next-hops. Even if some are suboptimal, there will be optimal alternatives that have been identified as well.

The number of 2.5 route errors seems small, but for many routes there is no suboptimal ``best-next-hop''. For example, any route from the top left node to any node in the bottom right half of the grid, will pass through either of the two neighbors of the top left node. Both neighbors are optimal choices.

The changes in the alternative model suggest that there may be some overhead since it sends more OGMs from more best-next-hops. On the other hand the alternative model will only rebroadcast duplicates if they improve on the TTL, unlike the literal model. We see that the maximum over 100 runs, rises for both models up to 45 OGMs in the buffer. Over all 255 time units the average number of OGMs in the buffer of a node in the literal model is 1.7, while it is 2.0 for the alternative model. It appears that the alternative model introduces a noticeable, but small overhead.

\section{Conclusion}
This paper modelled on a network routing protocol known as Better Approach to Mobile Ad hoc Networks
(B.A.T.M.A.N.).  The formalisation of the
protocol revealed several ambiguities and inconsistencies in the RFC. We developed two models, one implementing a literal reading of the RFC, the second model  closer to the underlying concepts. The literal model e.g implemented the requirement that there can only be one best-next-hop, whereas the alternative model relaxed the condition to include all nodes with the same ranking. On the other hand, the alternative model does rebroadcast fewer OGMs with a suboptimal TLL, i.e. OGMs that  arrived via suboptimal routes.

For a basic model we showed that both interpretations ensure loop-freedom, bidirectional link
discovery, and route-discovery. This served mostly to debug the model. We then use simulation of a timed model to compare the performance of the literal and alternative model, and found  that the alternative model identifies
a significantly lower number of suboptimal routes, and thus improves on the literal interpretation of
the RFC.

Future work should consider mobility in networks and check whether we can observe route discovery failures as has been predicted by Cigno and Furlan \cite{cignoimproving}. It would be interesting to see how the proposed alternative could be combined by suggestion for improvements made by these authors and others.

In general the process of formalisation has shown that it is important for an RFC to define all concepts, even those that seem obvious, such as ``newer sequence number''. It also shows that the RFC should clearly distinguish between the preconditions -- \emph{Rule 5.4} to \emph{Rule 5.7} in this paper -- and the processes itself. If  a process formulates those conditions again, even if it is only meant as a clarification, it can be a source of inconsistencies and ambiguities.

\vfill

%% file: main.bbl
\begin{thebibliography}{10}
\providecommand{\bibitemdeclare}[2]{}
\providecommand{\surnamestart}{}
\providecommand{\surnameend}{}
\providecommand{\urlprefix}{Available at }
\providecommand{\url}[1]{\texttt{#1}}
\providecommand{\href}[2]{\texttt{#2}}
\providecommand{\urlalt}[2]{\href{#1}{#2}}
\providecommand{\doi}[1]{doi:\urlalt{http://dx.doi.org/#1}{#1}}
\providecommand{\bibinfo}[2]{#2}

\bibitemdeclare{article}{bulychev2012uppaal}
\bibitem{bulychev2012uppaal}
\bibinfo{author}{Peter \surnamestart Bulychev\surnameend},
  \bibinfo{author}{Alexandre \surnamestart David\surnameend},
  \bibinfo{author}{Kim~Gulstrand \surnamestart Larsen\surnameend},
  \bibinfo{author}{Marius \surnamestart Miku{\v{c}}ionis\surnameend},
  \bibinfo{author}{Danny~B{\o}gsted \surnamestart Poulsen\surnameend},
  \bibinfo{author}{Axel \surnamestart Legay\surnameend} \&
  \bibinfo{author}{Zheng \surnamestart Wang\surnameend} (\bibinfo{year}{2012}):
  \emph{\bibinfo{title}{UPPAAL-SMC: Statistical model checking for priced timed
  automata}}.
\newblock {\sl \bibinfo{journal}{arXiv preprint arXiv:1207.1272}},
  \doi{10.4204/EPTCS.85.1}.

\bibitemdeclare{phdthesis}{cignoimproving}
\bibitem{cignoimproving}
\bibinfo{author}{Renato \surnamestart Cigno\surnameend} \&
  \bibinfo{author}{Daniele \surnamestart Furlan\surnameend}:
  \emph{\bibinfo{title}{Improving BATMAN Routing Stability and Performance}}.
\newblock Ph.D. thesis, \bibinfo{school}{PhD thesis, University of Trento,
  2011.}

\bibitemdeclare{article}{OLSR}
\bibitem{OLSR}
\bibinfo{author}{T.~\surnamestart Clausen\surnameend} \&
  \bibinfo{author}{Jacquet \surnamestart P.\surnameend}:
  \emph{\bibinfo{title}{{Optimized Link State Routing Protocol (OLSR), Network
  Working Group}}}.
\newblock \urlprefix\url{http://www.tools.ietf.org/html/rfc3626}.
\newblock \bibinfo{note}{Accessed:2015-06-08}.

\bibitemdeclare{inproceedings}{fehnker2013topology}
\bibitem{fehnker2013topology}
\bibinfo{author}{Ansgar \surnamestart Fehnker\surnameend},
  \bibinfo{author}{Peter \surnamestart H{\"o}fner\surnameend},
  \bibinfo{author}{Maryam \surnamestart Kamali\surnameend} \&
  \bibinfo{author}{Vinay \surnamestart Mehta\surnameend}
  (\bibinfo{year}{2013}): \emph{\bibinfo{title}{Topology-based mobility models
  for wireless networks}}.
\newblock In: {\sl \bibinfo{booktitle}{International Conference on Quantitative
  Evaluation of Systems}}, \bibinfo{organization}{Springer}, pp.
  \bibinfo{pages}{389--404}, \doi{10.1007/978-3-642-40196-1_32}.

\bibitemdeclare{inproceedings}{fehnker2012automated}
\bibitem{fehnker2012automated}
\bibinfo{author}{Ansgar \surnamestart Fehnker\surnameend}, \bibinfo{author}{Rob
  \surnamestart Van~Glabbeek\surnameend}, \bibinfo{author}{Peter \surnamestart
  H{\"o}fner\surnameend}, \bibinfo{author}{Annabelle \surnamestart
  McIver\surnameend}, \bibinfo{author}{Marius \surnamestart
  Portmann\surnameend} \& \bibinfo{author}{Wee~Lum \surnamestart
  Tan\surnameend} (\bibinfo{year}{2012}): \emph{\bibinfo{title}{Automated
  analysis of AODV using UPPAAL}}.
\newblock In: {\sl \bibinfo{booktitle}{International Conference on Tools and
  Algorithms for the Construction and Analysis of Systems}},
  \bibinfo{organization}{Springer}, pp. \bibinfo{pages}{173--187},
  \doi{10.1007/978-3-642-28756-5_13}.

\bibitemdeclare{article}{furlan2011analysis}
\bibitem{furlan2011analysis}
\bibinfo{author}{Daniele \surnamestart Furlan\surnameend}
  (\bibinfo{year}{2011}): \emph{\bibinfo{title}{Analysis of the overhead of
  BATMAN routing protocol in regular torus topologies}}.
\newblock {\sl \bibinfo{journal}{University of Trento, Italy, Tech. Rep}}.
\newblock
  \urlprefix\url{https://downloads.open-mesh.org/batman/papers/OGMoverhead.pdf}.

\bibitemdeclare{phdthesis}{paderborn}
\bibitem{paderborn}
\bibinfo{author}{Tobias \surnamestart Hardes\surnameend}
  (\bibinfo{year}{2015}): \emph{\bibinfo{title}{{Performance Analysis and
  Simulation of a Freifunk Mesh Network in Paderborn using B.A.T.M.A.N
  Advanced}}}.
\newblock \bibinfo{type}{Master's thesis}, \bibinfo{school}{University of
  Paderborn}.
\newblock
  \urlprefix\url{http://thardes.de/wp-content/uploads/2016/03/thesis.pdf}.

\bibitemdeclare{misc}{Huhtonen04comparingaodv}
\bibitem{Huhtonen04comparingaodv}
\bibinfo{author}{Aleksandr \surnamestart Huhtonen\surnameend}
  (\bibinfo{year}{2004}): \emph{\bibinfo{title}{Comparing AODV and OLSR Routing
  Protocols}}.
\newblock
  \urlprefix\url{http://www.tml.tkk.fi/Studies/T-110.551/2004/papers/Huhtonen.pdf}.

\bibitemdeclare{inproceedings}{Kamali2015}
\bibitem{Kamali2015}
\bibinfo{author}{Mojgan \surnamestart Kamali\surnameend},
  \bibinfo{author}{Peter \surnamestart H{\"o}fner\surnameend},
  \bibinfo{author}{Maryam \surnamestart Kamali\surnameend} \&
  \bibinfo{author}{Luigia \surnamestart Petre\surnameend}
  (\bibinfo{year}{2015}): \emph{\bibinfo{title}{Formal Analysis of Proactive,
  Distributed Routing}}.
\newblock In: {\sl \bibinfo{booktitle}{Software Engineering and Formal Methods:
  13th International Conference, SEFM 2015, York, UK, September 7-11, 2015.
  Proceedings}}, \bibinfo{publisher}{Springer}, pp. \bibinfo{pages}{175--189},
  \doi{10.1007/978-3-319-22969-0_13}.

\bibitemdeclare{article}{stairs}
\bibitem{stairs}
\bibinfo{author}{Elis \surnamestart Kulla\surnameend},
  \bibinfo{author}{Masahiro \surnamestart Hiyama\surnameend},
  \bibinfo{author}{Makoto \surnamestart Ikeda\surnameend} \&
  \bibinfo{author}{Leonard \surnamestart Barolli\surnameend}
  (\bibinfo{year}{2012}): \emph{\bibinfo{title}{Performance Comparison of OLSR
  and BATMAN Routing Protocols by a MANET Testbed in Stairs Environment}}.
\newblock {\sl \bibinfo{journal}{Comput. Math. Appl.}}
  \bibinfo{volume}{63}(\bibinfo{number}{2}), pp. \bibinfo{pages}{339--349},
  \doi{10.1016/j.camwa.2011.07.035}.

\bibitemdeclare{misc}{batman4omnet}
\bibitem{batman4omnet}
\bibinfo{author}{Spyridon \surnamestart Marinis~Artelaris\surnameend}
  (\bibinfo{year}{2016}): \emph{\bibinfo{title}{Performance evaluation of
  routing protocols for Wireless Mesh Networks}}.
\newblock
  \urlprefix\url{http://lnu.diva-portal.org/smash/get/diva2:903013/FULLTEXT01.pdf}.

\bibitemdeclare{article}{BatmanRFC}
\bibitem{BatmanRFC}
\bibinfo{author}{Axel \surnamestart Neumann\surnameend},
  \bibinfo{author}{Corinna \surnamestart Aichele\surnameend},
  \bibinfo{author}{Marek \surnamestart Lindner\surnameend} \&
  \bibinfo{author}{Simon \surnamestart Wunderlich\surnameend}
  (\bibinfo{year}{2008}): \emph{\bibinfo{title}{Better approach to mobile
  ad-hoc networking (BATMAN)}}.
\newblock {\sl \bibinfo{journal}{IETF draft}}, pp. \bibinfo{pages}{1--24}.
\newblock
  \urlprefix\url{https://tools.ietf.org/html/draft-wunderlich-openmesh-manet-routing-00}.

\bibitemdeclare{inproceedings}{van2013sequence}
\bibitem{van2013sequence}
\bibinfo{author}{Rob \surnamestart Van~Glabbeek\surnameend},
  \bibinfo{author}{Peter \surnamestart H{\"o}fner\surnameend},
  \bibinfo{author}{Wee~Lum \surnamestart Tan\surnameend} \&
  \bibinfo{author}{Marius \surnamestart Portmann\surnameend}
  (\bibinfo{year}{2013}): \emph{\bibinfo{title}{Sequence numbers do not
  guarantee loop freedom: AODV can yield routing loops}}.
\newblock In: {\sl \bibinfo{booktitle}{Proceedings of the 16th ACM
  international conference on Modeling, analysis \& simulation of wireless and
  mobile systems}}, \bibinfo{organization}{ACM}, pp. \bibinfo{pages}{91--100},
  \doi{10.1145/2507924.2507943}.

\bibitemdeclare{inproceedings}{hallway}
\bibitem{hallway}
\bibinfo{author}{J.~C.~P. \surnamestart Wang\surnameend},
  \bibinfo{author}{B.~\surnamestart Hagelstein\surnameend} \&
  \bibinfo{author}{M.~\surnamestart Abolhasan\surnameend}
  (\bibinfo{year}{2010}): \emph{\bibinfo{title}{Experimental evaluation of IEEE
  802.11s path selection protocols in a mesh testbed}}.
\newblock In: {\sl \bibinfo{booktitle}{2010 4th International Conference on
  Signal Processing and Communication Systems}}, pp. \bibinfo{pages}{1--3},
  \doi{10.1109/ICSPCS.2010.5709664}.

\bibitemdeclare{inproceedings}{wibling2004automatized}
\bibitem{wibling2004automatized}
\bibinfo{author}{Oskar \surnamestart Wibling\surnameend},
  \bibinfo{author}{Joachim \surnamestart Parrow\surnameend} \&
  \bibinfo{author}{Arnold \surnamestart Pears\surnameend}
  (\bibinfo{year}{2004}): \emph{\bibinfo{title}{Automatized verification of ad
  hoc routing protocols}}.
\newblock In: {\sl \bibinfo{booktitle}{International Conference on Formal
  Techniques for Networked and Distributed Systems}},
  \bibinfo{organization}{Springer}, pp. \bibinfo{pages}{343--358},
  \doi{10.1007/978-3-540-30232-2_22}.

\end{thebibliography}
